# Local transport measurements in laser patterned BSCCO-Ag tapes

C. F. Sánchez Valdés, C. Pérez-Penichet, C. Noda

Superconductivity Laboratory, Magnetism Laboratory, IMRE-Physics Faculty,

University of Havana, 10400 Havana, Cuba

M. Arronte

Technological Laser Laboratory, IMRE-Physics Faculty, University of Havana, 10400

Havana, Cuba

A. J. Batista-Leyva

Department of General Physics and Mathematics, Instec, 10400 Havana, Cuba

Ø. Haugen and T. H. Johansen

Department of Physics, University of Oslo, Blindern, N-0316 Oslo, Norway

Z. Han

Applied Superconductivity Research Center, Department of Physics, Tsinghua University

Beijing 100084, P. R. China

E. Altshuler

Superconductivity Laboratory, Magnetism Laboratory

and "Henri Poincaré" Group of Complex Systems,

IMRE-Physics Faculty, University of Havana, 10400 Havana, Cuba

The determination of inter- and intra-filament transport characteristics in superconducting composites such as BSCCO-Ag tapes is of great importance for material evaluation towards applications. Most attempts to separate the two contributions have relied on indirect methods based on magnetic measurements such as SQUID or magneto-optic imaging techniques. Here we show that laser patterning of superconducting BSCCO-Ag tapes constitutes a simple approach to measure local transport properties in a direct way, even able to separate inter- and intra-filament contributions to the overall transport behavior of the sample. Our technique is potentially useful for sub-millimeter devices based on superconducting tapes.

74.25.Fy, 74.25.Sv, 74.72.Hs, 74.81.Bd, 42.62.-b

One of the most difficult goals in the measurement of superconducting properties of inhomogeneous samples such as ceramic materials, polycrystalline films or multifilamentry tapes is to separate the contributions of intra- and inter-grain currents to the overall behavior of the samples. This is a fundamental issue for materials optimization, since there is no trivial correspondence between the microstructure and the connectivity in most samples. When macroscopic methods are used to separate intra- and inter-granular contributions, theoretical models are involved, which typically imply strong assumptions and approximations when interpreting results from magnetometric techniques such as SQUID [1,2] and conventional transport measurements [3,4]. Even more "local" magnetic sensing techniques such as micro Hall probes and magneto-optical imaging need the use of an inversion scheme to calculate the local currents in the sample [5,6]. A number of methods have been created to measure transport properties on different "stripes" along the axis of the tape: mechanical cuts [7,8], the application of "local" strong magnetic fields [9,10,11] and, in one case, the use of a photolithographic technique [12]. A very accurate method to study *isolated* filaments from BSCCO-Ag tapes, is to extract a filament by dissolving the silver matrix, trimming the filament crossection with a laser, and then perform the desired transport measurements [13]. It is worth noting that laser micromachining was recently applied also to YBCO coated superconductors as a way to fabricate microwave devices [14].

In this paper, we propose a new method to measure directly the transport properties at different locations and along different directions of a multifilamentary BSCCO-Ag tape using the conventional four-probe technique on samples locally patterned using a laser technique. The patterning is designed in such a way that local measurements of parameters such as the critical current could be performed along superconducting filaments (probing intra-filament properties) and transverse to the filaments (probing inter-filament properties). The transport measurements along the filament could even be done at different locations along the lateral dimension of the tapes.

Tapes of $Bi_2Sr_2Ca_2Cu_3O_{10+x}$-Ag (BSCCO-Ag) were fabricated by the powder-in-tube technique with subsequent drawing and rolling[15]. The tapes were 4.32 mm wide and 0.23 mm thick (0.22 mm without including Ag sheath), and contained 61 filaments. Each filament was 0.1-0.2 mm wide and a few microns thick. The overall long length critical current of the tape was 65 A at 77 K, corresponding to an engineering critical current density of 6.6 kA/cm$^2$.

Pieces of tape were patterned using a 1Nd:YAG PLC-Cut pulsed laser, with free running regime fundamental harmonic of 1064 nm, which produced a cutting spot on the sample through the optical system shown in Fig. 1. A suitable beam expander and an achromatic focusing lens with a focal length of 100 nm were used to control the spot size. The samples were glued on pieces of circuit board with a hole under the region to be laser-patterned, and then mounted on a x-y micro-stepping stage that could be moved along the focal plane of the lens. Parameters were adjusted empirically to minimize the formation of debris (solidified molten material around the processed area), leading us to use a pulse duration and energy of 130 µs and 300 mJ, respectively, applied with a repetition rate of 2 to 3 Hz. Then, the straight cuts were obtained as a chain of holes of 90 µm diameter, with 70% overlap between one hole and the next. The overall speed for straight cuts was 5 mm/min.

Transport measurements were performed using the four-probe technique with Ag-paste contacts. The temperature was controlled within the range 80 – 150 K with a resolution of ± 0.02 K using a Lake Shore 330 temperature controller with a Lake Shore DT470 silicon diode as thermometer. In the resistivity measurements, the sample was excited with 1 mA at 1 kHz. The voltage was amplified a thousand times with a low noise amplifier, and the resulting signal measured using a Scitec Instruments 500 MC lock in amplifier, with an output resolution better than 100 nV. We define the critical temperature, $T_c$, as the temperature value where the resistance departs from the low-temperature horizontal noise. The critical current was measured using a pulsed-current technique [16] with a voltage criterium of 1µV, and the engineering critical current density, $J_e$, was defined by dividing the critical current by the total crossection perpendicular to the current flow, which is the one relevant for transport applications.

Fig. 2a shows a piece of tape longitudinally sliced with the laser technique in such a way that three bridges were obtained. Two of the bridges, the one at the center of the tape (bridge I), and a second one near the edge of the tape (bridge II) were studied here. Fig. 2b shows a schematic of those bridges, indicating also where the images in Fig. 2c were taken. Fig. 3a shows a piece of tape sliced in such a way that a transversal bridge is obtained (bridge III). Fig. 3b shows a schematic of this bridge, indicating also where the image in Fig. 3c was taken. Notice in Figs. 2c and 3c that there are no visible effects of the laser heating on the superconducting filaments. A further transversal bridge of 1 mm width (bridge IIIwide) not shown in Fig. 3 was also studied. Notice that, in Fig. 2c, the transport current enters the picture perpendicular to the paper (or vice versa), while, in Fig. 3c, it runs from left to right (or vice versa).

Fig. 4 shows the resistive transition for bridges I, II, III, IIIwide and also the original tape. The curves have been normalized to the resistivity of bridge I at $T = 115$ K. While critical temperatures have moderately decreased after the laser cutting, the maximum decrease is around 1.5 K, and occurs for bridge III. However, the observed depression in $T_c$ is not exclusively connected to the laser heating effect, but also to the local microstructure "selected" in each bridge. For example, bridge I shows the smallest decrease in $T_c$, since it runs through the center of the tape, where filaments are probably quite uniform, as suggested by the magneto-optical images presented by Cai and co-workers [13]. Our bridge II shows a bigger shift, which is expected because it runs near the edge of the tape, where "unhealed cracks" are probably common, also suggested by magneto-optical pictures reported by Cai and co-workers[13]. Bridge III displays the smallest $T_c$ value, which may be related to the fact that the current has been forced to move transversally from one filament to another. While conduction through bridge IIIwide also goes transverse to the filaments, the bridge itself is wider, so the region affected by laser heating comprises a smaller proportion of the bridge's crossection perpendicular to the current flow. All in all, the decrease in $T_c$ in these cases is quite small, and, if necessary, could be minimized using an appropriate heat treatment. Moreover, the width of the resistive transitions is very similar for all measurements (nearly 1 K). Both facts indicate that our laser cutting technique affects very little the superconducting properties of the tape.

Figure 5 shows the details of the $J_e$ vs. $T$ curves for bridges I and III near $T_c$ (absolute errors can be taken as ± 100 A/cm$^2$ for the experimental points shown in the figure). Notice that, for bridge I, we have calculated $J_e$ using the total crossection in the plane of the paper as shown in Fig. 2c, while for bridge III, we have used the total crossection perpendicular to the plane of the paper, which is not the one shown in Fig. 3c. As expected, the engineering critical current density for the transversal bridge, III, is smaller than that for the longitudinal bridge I, since in the former geometry, the transport current is forced to cross through different filaments. Moreover, the linear behavior of the $J_e(T)$ curve shown in Fig. 5 from bridge I is typical of good BSCCO tapes measured along their long axis [17], while the curvature in the $J_e(T)$ curve for bridge III can be interpreted as a fingerprint of SIS or SNS behavior[18] commonly observed in the inter-granular critical currents of High $T_c$ polycrystals in the presence of a distribution of junction qualities [19]. Finally, it must be pointed out that our measurements do not show the "crossover" behaviour in the intra- and inter-grain $I_c(T)$ curves detected by Bobyl *et al.* for BSCCO tapes through an inversion scheme from magneto-optical experiments [6]. The engineering critical current densities at 77 K for bridges I and III where 6500 ± 500 A/cm$^2$ and 3500 ± 500 A/cm$^2$, respectively.

In summary, we have devised a laser technique that allows performing local transport measurements in regions and directions selected *ad hoc* within a BSCCO-Ag tape, which constitutes a new way for the characterization of superconducting tapes and related materials. It is also potentially useful for the construction of sub-millimeter-sized devices based on superconducting tapes.

Figure captions

**Figure 1 Schematic of the laser cutting technique.**

**Figure 2 Longitudinal bridges cut on a BSCCO tape using a laser technique.**

**Figure 3 Transversal bridge cut on a BSCCO tape using a laser technique**

**Figure 4 Resistance vs. temperature curves for the non patterned tape, and for bridges I, II, III and IIIwide. In all cases, the resistance has been normalized to that corresponding to bridge I at 115 K.**

**Figure 5 Engineering critical current density vs. temperature for bridges I and III.**

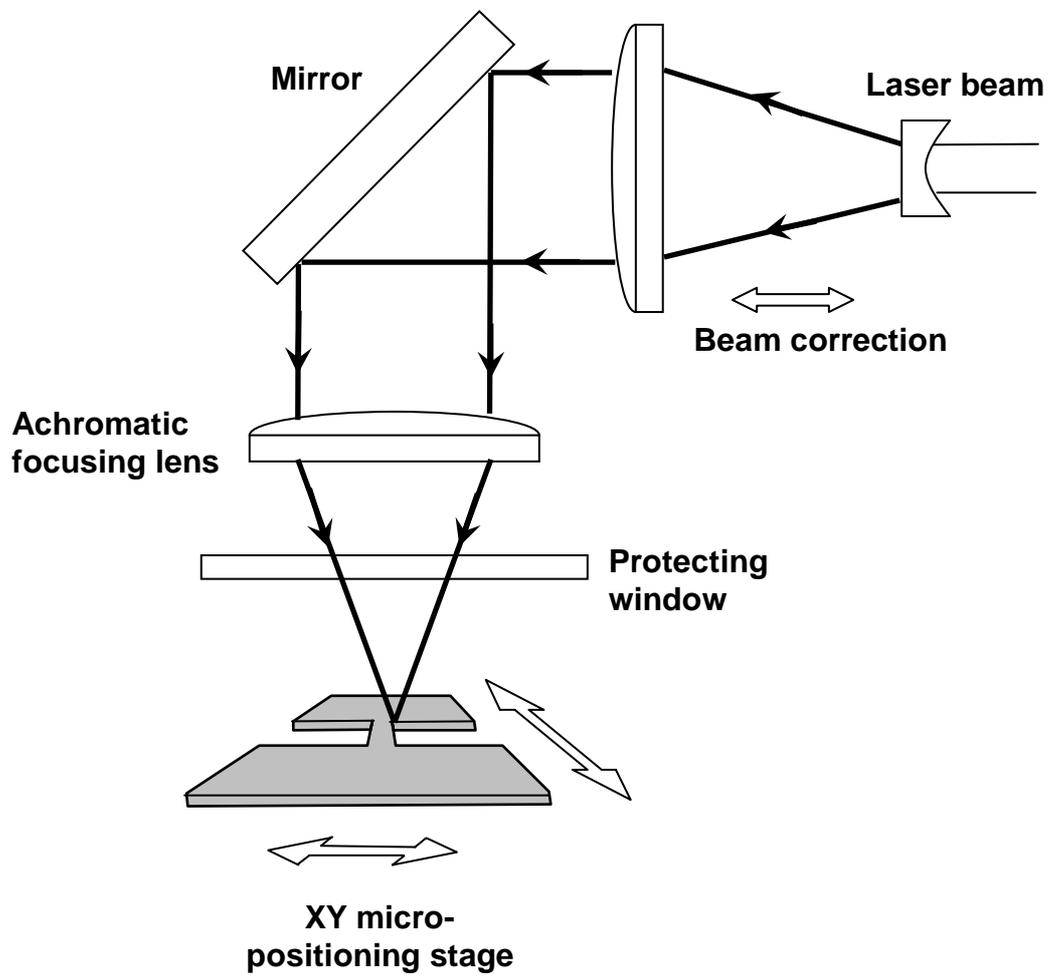

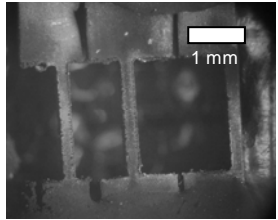

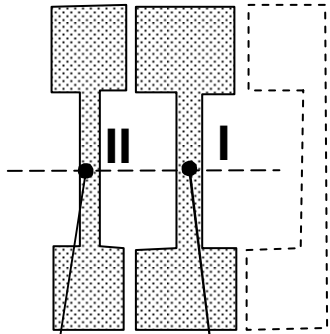

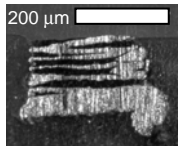 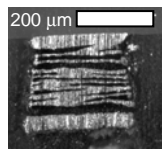

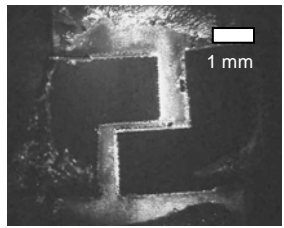

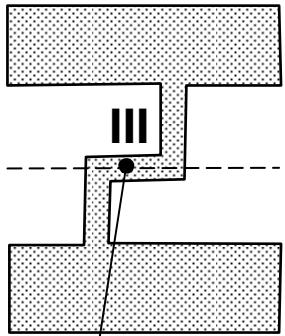

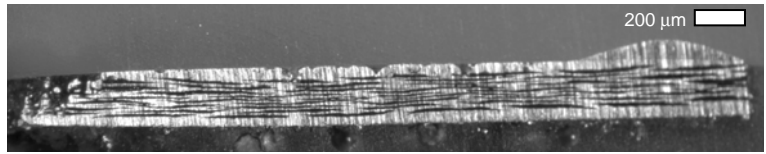

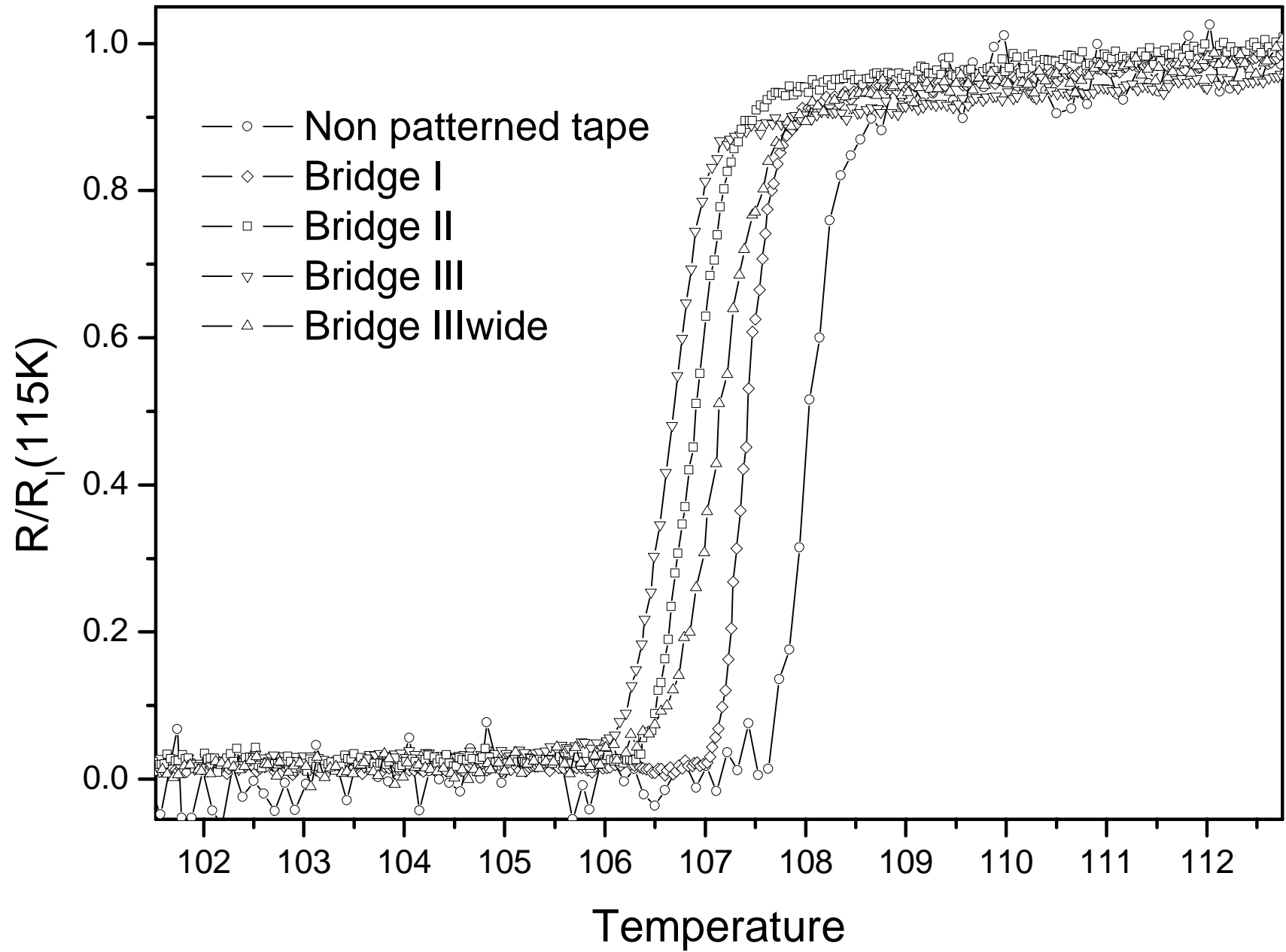

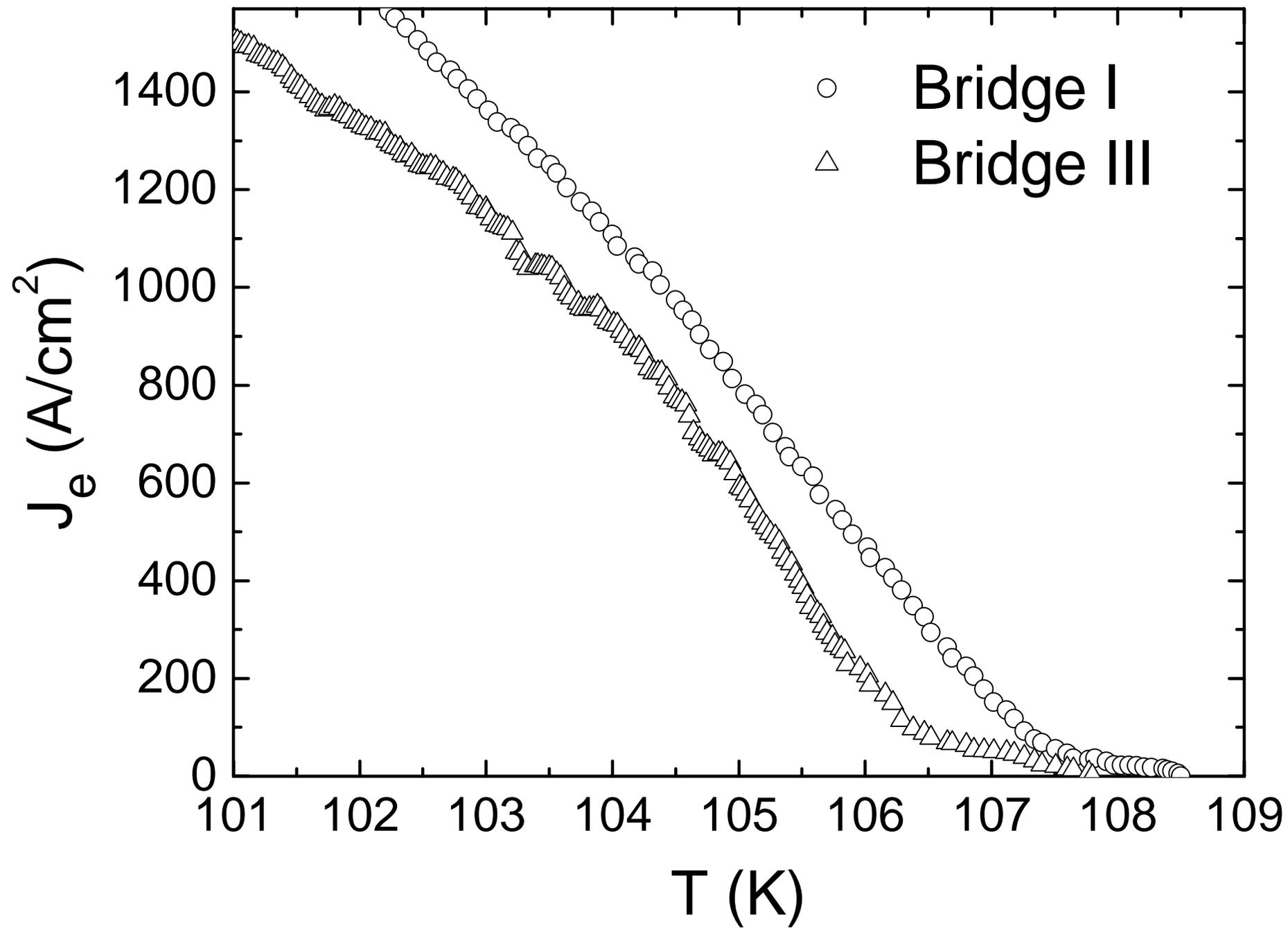